\documentclass[11pt]{article}

\usepackage[T1]{fontenc}
\usepackage[utf8]{inputenc}

\usepackage{amsmath,amssymb,amsthm}
\usepackage{mathtools}
\usepackage{geometry}
\usepackage{hyperref}
\usepackage{array}

\newcommand{\Tau}{\mathcal{T}}

\newcommand{\detz}[1]{\mathop{\mathrm{det}}\nolimits_{\zeta}\!\left(#1\right)}

\title{\textbf{
A Differential Geometry and Algebraic Topology Based\\
Public-Key Cryptographic Algorithm\\
in the Presence of Quantum Adversaries
}}

\author{Andrea Rondelli}
\date{}

\begin{document}
\maketitle

\begin{abstract}
\emph{
In antiquity, the seal embodied trust, secrecy, and integrity in safeguarding the exchange of letters and messages.
}
The purpose of this work is to continue this tradition in the contemporary era,
characterized by the presence of quantum computers,
classical supercomputers,
and increasingly sophisticated artificial intelligence.

We introduce \textbf{Z-Sigil}, an asymmetric public-key cryptographic algorithm grounded in
functional analysis,
differential geometry,
and algebraic topology,
with the explicit goal of achieving resistance against both classical and quantum attacks.
The construction operates over the tangent fiber bundle of a compact Calabi--Yau manifold
\cite{GreenSchwarzWitten},
where cryptographic keys are elements of vector tangent fibers,
with a binary operation defined on tangent spaces \(T_p(\mathcal{M})\) giving rise to a groupoid structure.

Encryption and decryption are performed iteratively on message blocks,
enforcing a serial architecture designed to limit quantum parallelism
\cite{Lloyd2000,Preskill}.
Each block depends on secret geometric and analytic data,
including a randomly chosen base point on the manifold,
a selected section of the tangent fiber bundle,
and auxiliary analytic data derived from operator determinants
and Zeta function regularization
\cite{Hawking}.

The correctness and invertibility of the proposed algorithm are proven analytically.
Furthermore,
any adversarial attempt to recover the plaintext without the private key leads to an exponential growth
of the adversarial search space,
even under quantum speedups.
The use of continuous geometric structures,
non-linear operator compositions,
and enforced blockwise serialization distinguishes this approach from existing quantum-safe
cryptographic proposals based on primary discrete algebraic assumptions.
\end{abstract}

\section{Introduction}

Modern cryptography rests on three fundamental pillars:
confidentiality,
integrity,
and authenticity.
Public-key cryptography enables secure communication over untrusted channels
by allowing encryption keys to be transmitted openly,
while keeping decryption keys private
\cite{BonehShoup}.
Classical constructions such as RSA and Diffie--Hellman rely on arithmetic problems commonly regarded
as computationally intractable for classical computers
\cite{Shoup};
however, the advent of quantum computing poses a serious threat to these assumptions
\cite{Shor,NielsenChuang,Watrous}.

Quantum algorithms such as Shor’s factoring algorithm and Grover’s search algorithm
significantly reduce the complexity of problems underpinning widely deployed cryptographic systems
\cite{Shor,Grover}.
As a result, there is growing interest in constructions whose security does not rely solely on
number-theoretic hardness, but instead exploits richer mathematical structures.

A central design choice in this work is the use of a blockwise, iterative encryption process.
The message is divided into components, and encryption proceeds sequentially throughout an overall cycle.
This architecture is intentionally introduced to limit quantum parallelism:
to decrypt block \(i\), an adversary must first compute the data associated with component \(i-1\).
This dependency introduces an explicit computational relation across components,
reducing naive parallel evaluation across components
\cite{Lloyd2000,Preskill}.

The remainder of the paper is organized as follows:
Section~2 introduces the mathematical background,
Section~3 defines the encryption and decryption maps,
Section~4 proves invertibility and integrity,
Section~5 analyzes computational complexity and quantum security,
Section~6 presents a high-level structural analogy with RSA,
and conclusions are discussed in Section~7.

\section{Mathematical Setting and Preliminaries}

\subsection{Kähler and Calabi--Yau manifolds}

\textbf{Definition 1.}
A Kähler manifold \((\mathcal{M},g,J,\omega)\) is a smooth complex manifold endowed with a Hermitian metric \(g\),
a complex structure \(J\),
and a closed Kähler form \(\omega\) satisfying
\[
\omega(X,Y)=g(JX,Y), \qquad d\omega=0,
\]
where \(d\) denotes the exterior derivative.

\textbf{Definition 2.}
A Calabi--Yau manifold is a compact Kähler manifold with trivial canonical bundle,
equivalently admitting a Ricci-flat Kähler metric.
An equivalent characterization is that the first Chern class satisfies \(c_1(\mathcal{M})=0\).

Calabi--Yau manifolds play a central role in string theory as models for compactified extra dimensions,
historically rooted in dual resonance models and the Veneziano amplitude
\cite{Veneziano}.
In modern formulations, fundamental particles arise as vibrational modes of one-dimensional strings,
and higher-dimensional extended objects (D-branes) generalize this picture
\cite{Zwiebach,GreenSchwarzWitten,Deligne}.
These geometric structures motivate the use of Calabi--Yau spaces as high-complexity continuous key spaces.
For complementary physical background, see also \cite{Kaku}.

Let \(r=\dim_{\mathbb{R}}\mathcal{M}\);
the complex dimension is \(r/2\),
treated as a model parameter to be determined via numerical simulations and optimization.

\subsection{Tangent bundles, fibers, and sections}

Let \(T\mathcal{M}\to\mathcal{M}\) denote the tangent fiber bundle of \(\mathcal{M}\).
For each point \(p\in\mathcal{M}\), the fiber \(T_p\mathcal{M}\) is a real vector space of dimension \(r\).

\textbf{Definition 3.}
A (local) section of \(T\mathcal{M}\) is a smooth map
\[
\sigma:U\subset\mathcal{M}\to T\mathcal{M}
\]
such that \(\pi\circ\sigma=\mathrm{id}_U\),
where \(\pi:T\mathcal{M}\to\mathcal{M}\) is the bundle projection.

In this construction, cryptographic keys are elements of tangent fibers,
while sections encode hidden geometric choices unavailable to an adversary.

\subsection{Groupoid structure on fibers}

\textbf{Definition 4.}
A \emph{groupoid} is a category in which every morphism is invertible
\cite{Spivak,Riehl}.
Explicitly, \(\mathcal{G}\) consists of a set of objects \(\mathrm{Ob}(\mathcal{G})\),
a set of morphisms \(\mathrm{Mor}(\mathcal{G})\),
source and target maps \(s,t:\mathrm{Mor}(\mathcal{G})\to\mathrm{Ob}(\mathcal{G})\),
identity morphisms for each object,
and a partially defined composition that is associative whenever defined.

For each \(p\in\mathcal{M}\), we define a fiberwise operation
\[
\circledast_p:T_p\mathcal{M}\times T_p\mathcal{M}\to\mathrm{End}(T_p\mathcal{M}),
\]
which is associative whenever defined, but not necessarily commutative.
In concrete realizations, \(\circledast_p\) is encoded by an \(r\times r\) matrix whose coefficients are non-linear
functions of local coordinates and depend explicitly on the base point \(p\).
The base point \(p_i\) is chosen uniformly at random and independently for each encryption block,
so that both the tangent fiber and the induced operation vary from block to block.

\textbf{Assumption 1.}
For each \(p\in\mathcal{M}\),
there exists an identity operator \(\mathbf{1}_{r\times r}\) such that for every public key
\(e\in T_p\mathcal{M}\) there exists a unique private key \(d\in T_p\mathcal{M}\) satisfying
\[
e\circledast_p d=\mathbf{1}_{r\times r}.
\tag{1}
\]
Since the inverse is well defined and unique, the induced correspondence on \(T_p\mathcal{M}\) is bijective.

\subsection{Gaussian Unitary Ensemble matrices}

\textbf{Definition 5.}
A random matrix \(G\in\mathbb{C}^{n\times n}\) is said to be drawn from the Gaussian Unitary Ensemble (GUE)
if it is Hermitian and its entries are distributed according to the standard GUE measure.
In practical implementations we enforce \(\det(G)\neq0\).

\subsection{Hilbert--Schmidt operators and zeta regularization}

Let \(\mathcal{H}\) be a separable Hilbert space.

\textbf{Definition 6.}
An operator \(A:\mathcal{H}\to\mathcal{H}\) is Hilbert--Schmidt if
\[
\|A\|_{HS}^2=\sum_j\|Ae_j\|^2<\infty
\]
for some (hence any) orthonormal basis \(\{e_j\}\).
We further assume \(A\) to be compact, essentially self-adjoint, and of trace class,
with real eigenvalues \(\lambda_j\in\mathbb{R}\) satisfying \(\lambda_j\to0\) as \(j\to\infty\).

\textbf{Definition 7.}
The spectral zeta function of \(A\) is
\[
\zeta_A(s)=\sum_j\lambda_j^{-s},
\]
where \(s\in\mathbb{C}\),
defined initially on a right half-plane and extended by analytic continuation.
The zeta-regularized determinant is

\[
\detz{A}=\exp\!\bigl(-\zeta_A'(0)\bigr).
\]

In this context, we also recall the classical Riemann zeta function
\[
\zeta(s)=\sum_{n=1}^{\infty}n^{-s},
\qquad n\in\mathbb{N},\ \ s\in\mathbb{C},
\]
defined for \(\mathrm{Re}(s)>1\) and extended by analytic continuation.
The function admits an analytic continuation to the entire complex plane \(\mathbb{C}\),
holomorphic on \(\mathbb{C}\setminus\{1\}\), with a simple pole at \(s=1\).
Its nontrivial zeros are denoted
\[
\rho_k=\tfrac{1}{2}+i\gamma_k,
\]
and we use only the imaginary parts \(\gamma_k\neq0\).
The Riemann Hypothesis (still unproven) asserts that all nontrivial zeros satisfy \(\mathrm{Re}(\rho_k)=\tfrac12\).

\section{Encryption and Decryption Maps}

Like in the standard public-key communication setting,
Alice encodes a plaintext message as an integer string and partitions it into blocks
\[
m\mapsto(m_1,\dots,m_D),
\]
where \(D\) denotes the total number of blocks.

We define the encryption map
\[
\mathfrak{E}:\mathbb{Z}^+\to T_{p_i}\mathcal{M},
\]
and the decryption map
\[
\mathfrak{D}:T_{p_i}\mathcal{M}\to\mathbb{Z}^+.
\]

Encryption is given by
\[
c_i=\mathfrak{E}(m_i)=m_i\,N_{i-1}\,e_i,
\tag{2}
\]
where \(e_i\in T_{p_i}\mathcal{M}\) is the public key and
\[
N_i=\det(G_i)\,\detz{A_i}\,\prod_k\gamma_k,
\qquad N_i\neq0.
\tag{3}
\]

Let \(\Tau:\mathrm{End}(T_{p_i}\mathcal{M})\to\mathbb{R}\) denote the normalized trace functional
\[
\Tau(X)\coloneqq \frac{1}{r}\mathrm{Tr}(X).
\]

Decryption is defined by
\[
m_i=\mathfrak{D}(c_i)=
\Tau\!\left(\frac{1}{N_{i-1}}\bigl(c_i\circledast_{p_i}d_i\bigr)\right),
\tag{4}
\]
where \(d_i\in T_{p_i}\mathcal{M}\) is the private key satisfying Assumption~1.
By construction, the output lies in \(\mathbb{Z}^+\),
enabling the inverse UTF-16 decoding map to recover the original alphanumeric string.

\section{Invertibility and Integrity}

Message integrity means that the receiver recovers exactly the same plaintext message that was originally sent.

\textbf{Theorem 1.}
For every plaintext block \(m_i\in\mathbb{Z}^+\), we have
\[
(\mathfrak{D}\circ\mathfrak{E})(m_i)=m_i.
\]

\textbf{Proof.}
By definition,
\[
\mathcal{D}(\mathcal{E}(m_i))
=\tau\!\left(\frac{1}{N_i^{-1}}\,(m_i N_i^{-1} e_i \boldsymbol{\circledast}_{p_i} d_i)\right)
=\tau\!\left(m_i\,(e_i \boldsymbol{\circledast}_{p_i} d_i)\right)
=\tau(m_i\,\mathbf{1}_{r\times r})
= m_i\,\tau(\mathbf{1}_{r\times r})
= m_i.
\]
where we used Assumption~1, namely $e_i \boldsymbol{\circledast}_{p_i} d_i=\mathbf{1}_{r\times r}$, and the normalization
\[
\tau(\mathbf{1}_{r\times r})=\frac{1}{r}\mathrm{Tr}(\mathbf{1}_{r\times r})=1 \in \mathbb{R}.
\]
\textbf{QED.}

Thus the decryption map is a two-sided inverse of the encryption map. The correspondence between plaintext blocks and ciphertext blocks is therefore bijective, i.e., both injective and surjective.

\section{Computational Complexity and Quantum Security}

\subsection{Attacker model}

We consider a man-in-the-middle adversarial scenario in which an attacker intercepts the public
ciphertext--key pairs \((c_i,e_i)\) transmitted over an untrusted channel and attempts to recover the corresponding
private keys \(d_i\).
Each block depends on a hidden triple
\[
(p_i,\sigma_i,\circledast_{p_i}),
\]
chosen randomly and independently per block, together with the auxiliary multiplicative factor \(N_{i-1}\).

\subsection{Discretization and search space}

Let \(n\) denote the public-key bit-length.
Discretization induces a finite candidate space of size \(S(n)\).

\textbf{Assumption 2.}
From the adversary’s perspective, inversion behaves as an unstructured search over \(S(n)\).

\subsection{Oracle formulation}

The inversion task can be cast in the standard quantum query model.
Define an oracle that marks a candidate private key \(d_i\) if and only if it satisfies the inversion condition in (1);
for the public key \(e_i\), this means that
\[
e_i\circledast_{p_i} d_i=\mathbf{1}_{r\times r},
\]
together with consistency of the auxiliary multiplicative factor \(N_{i-1}\) in (2)--(4).
Since the hidden geometric data \((p_i,\sigma_i,\circledast_{p_i})\) varies blockwise and is unavailable to the attacker,
the oracle provides no exploitable algebraic structure beyond a yes/no decision.

\subsection{Grover lower bound and optimality}

\textbf{Theorem 2.}
If \(S(n)\ge2^{\alpha n}\) for some \(\alpha>0\),
then any Grover-type quantum attack requires \(\Omega(2^{\alpha n/2})\) oracle calls
and hence exponential gate complexity.

\textbf{Proof.}
In the quantum query model, Grover’s algorithm requires
\[
Q(n)=\Theta(\sqrt{S(n)})
\]
oracle calls.
If \(S(n)\ge 2^{\alpha n}\) for some \(\alpha>0\), then
\[
Q(n)=\Omega(\sqrt{S(n)})\ge \Omega\!\left(2^{\alpha n/2}\right).
\]
Let \(g(n)\) denote the gate cost of implementing one oracle call.
In any circuit model we have \(g(n)\ge 1\), and in concrete implementations we assume
\(g(n)\le \mathrm{poly}(n)\).
Therefore the total gate complexity satisfies
\[
G(n)=\Omega(Q(n)g(n))\ge \Omega\!\left(2^{\alpha n/2}\right),
\]
and under the standard assumption \(g(n)\le \mathrm{poly}(n)\) we obtain the sharper bound
\[
G(n)=\Omega\!\left(2^{\alpha n/2}\mathrm{poly}(n)\right).
\]
\textbf{QED.}

\textbf{Remark.}
For \(n=1024\) and \(\alpha=1\), the query lower bound is \(2^{512}\).
Using \(2^{512}=10^{512\log_{10}2}\), we obtain \(2^{512}\approx 10^{154}\).
This bound already exceeds by many orders of magnitude any physically meaningful brute-force regime,
including cosmological entropy estimates on the order of \(10^{120}\)–\(10^{122}\),
which quantify the maximum number of independent physical degrees of freedom available in the observable universe,
even under optimal Grover-type quantum speedups.

\subsection{Impact of blockwise serialization}

Because encryption is performed blockwise with an explicit dependency across successive components,
an adversary cannot evaluate multiple blocks independently in parallel.
This suppresses naive quantum parallelism across message blocks and prevents amplitude-amplification strategies
spanning multiple ciphertext components without first resolving the preceding ones.

\section{High-Level Structural Analogy with RSA}

At a very abstract level, both RSA and Z-Sigil rely on a public object enabling encryption and a hidden structure
enabling inversion.
In RSA, the public modulus \(N\) is constructed as the product of large primes, typically written as \(N=pq\),
while the security of the scheme relies on the computational hardness of recovering the prime factors from \(N\).
Inversion is controlled by a single algebraic secret \(\varphi(N)\).
In Z-Sigil, inversion requires access to the geometric core \((p_i,\sigma_i,\circledast_{p_i})\)
together with an additional analytic layer \(N_i\),
yielding a genuinely multilevel construction with no direct analogue in RSA.

\vspace{1em}
\begin{center}
\renewcommand{\arraystretch}{1.2}
\begin{tabular}{|>{\raggedright}p{4cm}|c|c|}
\hline
\textbf{Aspect} & \textbf{RSA} & \textbf{Z-Sigil} \\
\hline
Public object
&
$(N,e),\; N\in\mathbb{N}$
&
$e_i\in T_{p_i}\mathcal{M}$
\\
\hline
Core secret enabling inversion
&
$\varphi(N)$ (equiv.\ $p,q$)
&
$(p_i,\sigma_i,\circledast_{p_i})$
\\
\hline
Additional layers
&
None
&
$N_i=\det(G_i)\detz{A_i}\prod_k\gamma_k$
\\
\hline
Underlying structure
&
Algebraic
&
Geometric--topological
\\
\hline
\end{tabular}
\end{center}

\section{Conclusion}

We have proved correctness and integrity and have shown that key inversion reduces to an exponentially large
unstructured search, remaining quantum-safe even under Grover-type speedups.
The security of the construction derives from differential geometry and algebraic topology,
augmented by analytic operator-theoretic layers.

Future work will focus on extensive numerical testing and simulation to calibrate the remaining free parameters
of the construction, in particular the dimensional and geometric regimes of the underlying Calabi--Yau setting.
The goal is to identify an optimal operating point balancing computational efficiency and cryptographic security.
This includes implementation benchmarks on classical hardware and systematic simulations of both classical and
quantum attack models, with particular emphasis on Grover-type adversaries.

\end{document}